# Engineering of selective carrier injection in patterned arrays of single-quantum-dot entangled photon light-emitting diodes

T. H. Chung*, G. Juska*, S. T. Moroni, A. Pescaglini, A. Gocalinska and E. Pelucchi

*Tyndall National Institute, University College Cork, Lee Maltings, Cork, Ireland*

*T.H.C. and G.J. equally contributed to this work

## Abstract

**Scalability and foundry compatibility (as for example in conventional silicon based integrated computer processors) in developing quantum technologies are exceptional challenges facing current research. Here we introduce a quantum photonic technology potentially enabling large scale fabrication of semiconductor-based, site-controlled, scalable arrays of electrically driven sources of polarization-entangled photons, with the potential to encode quantum information. The design of the sources is based on quantum dots grown in micron-sized pyramidal recesses along the crystallographic direction (111)B theoretically ensuring high symmetry of the quantum dots – the condition for actual bright entangled photon emission. A selective electric injection scheme in these non-planar structures allows obtaining a high density of light-emitting diodes, with some producing entangled photon pairs also violating Bell's inequality. Compatibility with semiconductor fabrication technology, good reproducibility and control of the position make these devices attractive candidates for integrated photonic circuits for quantum information processing.**

To develop quantum technologies, the scientific community is looking into several alternative practical routes, varying as much as superconducting qubits, atoms on-chips, photonic integrated circuits and others[1,2,3,4]. All the explored technologies have to solve the scalability and reproducibility problem if they are to deliver successful real-life applications. In the case of the photonic quantum technologies, scalability requires moving from discrete optical elements to integrated photonic circuits and to on-chip solid-state sources, allowing, for example, thousands of units operating in unison – a condition which is very hard to fulfil at the moment.

Semiconductor quantum dot technology is fundamentally compatible with modern fab/foundry processes, and on-demand identical, single and entangled photons have been all demonstrated by optical pumping[5,6,7,8,9,10,11,12,13,14]. Nevertheless, while the development of electrically pumped (EP) quantum light sources has advanced in general[15], the development of a particular resource, EP entangled photon sources, has proven more challenging. After the first report in Nature[16], the community had to wait several years before a similar result could be obtained by other groups[17]. Importantly, the few devices reported to date, utilized epitaxial self-assembled QD structures, i.e. with no control on the source location, nor on the number of sources in a single device (typically hundreds or more, and not just one or, in the best case scenario, a few): a critical aspect for photonic integration scaling.

## Pyramidal quantum dot system

The technology presented herein is based on the Pyramidal QD system, recently highlighted for its inherent capability of delivering arrays of highly symmetric and uniform QDs[11,18]. The system is, nevertheless, intrinsically non planar, a feature which has impeded up to now the development of efficient electrically driven light emitting sources. Before discussing the relevant quantum optics results, we need to highlight the complexity of the Pyramidal system as a key ingredient. In short, single QDs are epitaxially grown by metalorganic vapour phase epitaxy (MOVPE) inside inverted pyramidal recesses, lithographically patterned on a (111)B GaAs substrate (a fragment of such template is shown in Fig. 1a). The structure comprises a number of differently composed III-V (Al)GaAs layers and an InGaAs QD layer (see Supplementary Material for a detailed description of each layer, and the reasons for inserting them) – all obeying complex epitaxial dynamics as reported elsewhere[19,20,21,22]. The outcome is an ensemble of self-forming nanostructures inside each pyramidal recess, described by the generic sketches in Fig. 1a. During growth of an AlGaAs alloy, fast diffusing Ga tends to segregate in the regions of intersecting walls of the recess and in a narrow region at the centre of the Pyramidal structure, effectively forming three embedded low bandgap vertical quantum wells (VQW) and a vertical quantum wire (VQWR) of ~20 nm in diameter, respectively. In addition, a thin InGaAs layer forms a group of interconnecting nanostructures: a flat quantum dot at the central axis of the structure, three lateral quantum wires (LQWR) and three lateral quantum wells (LQW). Such apex-down geometry hinders light extraction, which can be efficiently enhanced by two or three orders of magnitude (typically achieving $6 \cdot 10^4 - 15 \cdot 10^4$ photon/s detection rate under continuous-wave excitation in our system) selectively etching away the substrate (a process known as back-etching, see Methods) so to obtain apex-up pyramidal structures as shown in Fig. 1b – a typical configuration used for measurements[11,23]. To fabricate a single QD light-emitting diode, the lack of planarity on both sides of a sample, unfortunately, does not allow for simply contacting the back and the top of the Pyramid. The proposed schematic for a µLED device is depicted in Fig. 1c, emphasizing its 3D intrinsic nature, lateral dimensions ≪10µm, and drafting the presence of a nanowire like structure (VQWR) running through most of the structure, and, as it will be discussed later, contributing importantly to a selective injection process feeding the single QD at the centre (the magnified region is shown in Fig. 1d).

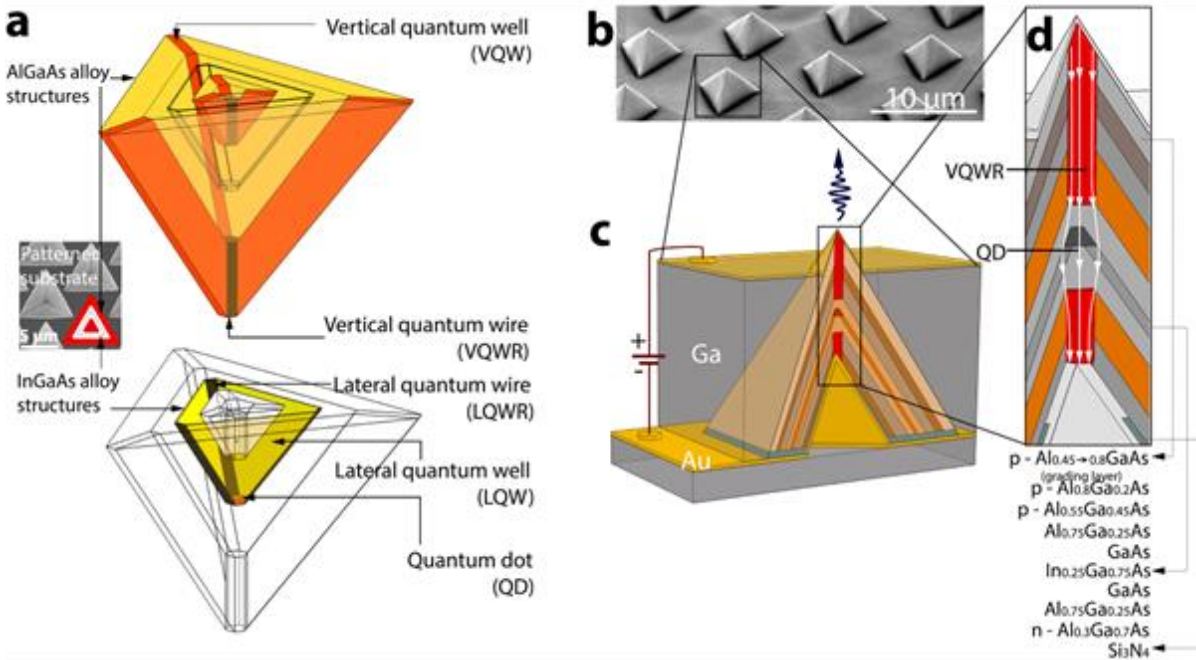

**Figure 1. The internal structure of a device and the schematics of a μLED. a**, The complex ensemble of nanostructures which self-forms within the pyramidal recesses pre-etched in a GaAs substrate. AlGaAs alloys forms gallium enriched structures: a vertical quantum wire (VQWR) along the central axis of the pyramid, and 3 vertical quantum wells (VQW). A nominally thin (0.5 nm) InGaAs alloy forms a QD, 3 lateral quantum wires (LQWR) and 3 lateral quantum wells (LQW). **b**, The SEM image of a sample right after the chemical etching step showing exposed pyramidal structures in an apex-up geometry which enhances light extraction by a few orders of magnitude. **c**, A sketch of a p-i-n junction μLED in cross-section view. **d**, A magnified region of the central part of a pyramid with a QD. The shown epitaxial layers comprise a representative structure with dominant AlGaAs alloys which form a vertical quantum wire (VQWR). Arrows indicate injection current through the region of the VQWR.

## Selective current injection

In order to fabricate a QD LED, there are number of hurdles to overcome, and we will here discuss the most relevant. Fig. 2a shows the cross-sectional SEM image of the Pyramidal structure, where the device is still contained inside the GaAs substrate (the different grey layers result from the various Al containing alloys and doping). For one thing, the centre vertical path (blue arrow in Fig. 2a), where the single QD is located, is not the most electrically favourable path. Because of the geometry, the centre path is ~3 times longer than taking a short across the side (red arrow Fig. 2a), i.e. in a LED structure current will favourably go along the red path, resulting in a low (or no) probability of intercepting the single QD, and, in general, in an overly crowded LED (i.e. too many charges running) impeding good quality emission from the QD, due to spectral wandering/decoherence induced by random electric fields fluctuations. Moreover, the system is even more non-ideal in its nature, presenting strong faceting at the top (this happens because there is virtually no growth on the planar (111B) substrate, resulting in "facet crowding" – see Supplementary Material and the region circled in Fig. 2a). These lateral facets are the regions of potential current shortcut.

To effectively achieve selective injection to the single QD, one needs to exploit an extra feature of the Pyramidal system – the formation of vertical quantum wires (VQWR) during the high bandgap $Al_{0.75}Ga_{0.25}As$ growth (Fig. 1a). This low bandgap ($Al_xGa_{1-x}As$, x≤0.3) embedded nanowire is expected to act as a low resistivity path for electrical charges to reach the InGaAs QD (i.e. to introduce a low resistivity path for charges at the centre only, while effectively leaving a high bandgap barrier for transport in the form of, e.g., the $Al_{0.75}Ga_{0.25}As$/ $Al_{0.3}Ga_{0.7}As$ interface, on the sides). To test our initial intuition, we performed finite elements simulations (see also Methods): the current density distribution within the magnified region with VQWRs of a simplified device model is shown in Fig. 2a. The simulation coherently describes a preferential current path at the centre. We also observe that the VQWR does not impede the current flow to other (quantum) structures, but nevertheless most of the current is restricted to the centre region. Based on the simulation results, we implemented this idea in a practical design of the device (Fig. 1c and 1d).

To force current injection at the centre of a Pyramid, a small Ti/Au contact has to be created in that region by a couple of processing steps. First, to prevent electrical short circuits outside the region of injection, a layer of $Si_3N_4$ is deposited onto the as-grown sample. Second, a tilted sample is evaporated 3 times by Ti/Au, rotating it by 120° in sample growth plane each time. The resulting metallic layer acts as a mask metal, so to leave a small aperture of $Si_3N_4$ at the centre of the recess (Fig. 2c). This is then opened by $CF_4$ plasma etching. Subsequently, a Ti/Au layer is evaporated as n-contact (and bonding) metal. The result is a small contact area (>100 $nm^2$) for the n/p region only at the centre of the Pyramid for the µLED, all simply exploiting self-aligning processes and no sophisticated high-resolution lithography. To prepare for the substrate removal the sample is bonded onto another Ti/Au-coated GaAs substrate with the assistance of a Sn-Au ribbon via a reflow soldering process, which provides a large area contact.

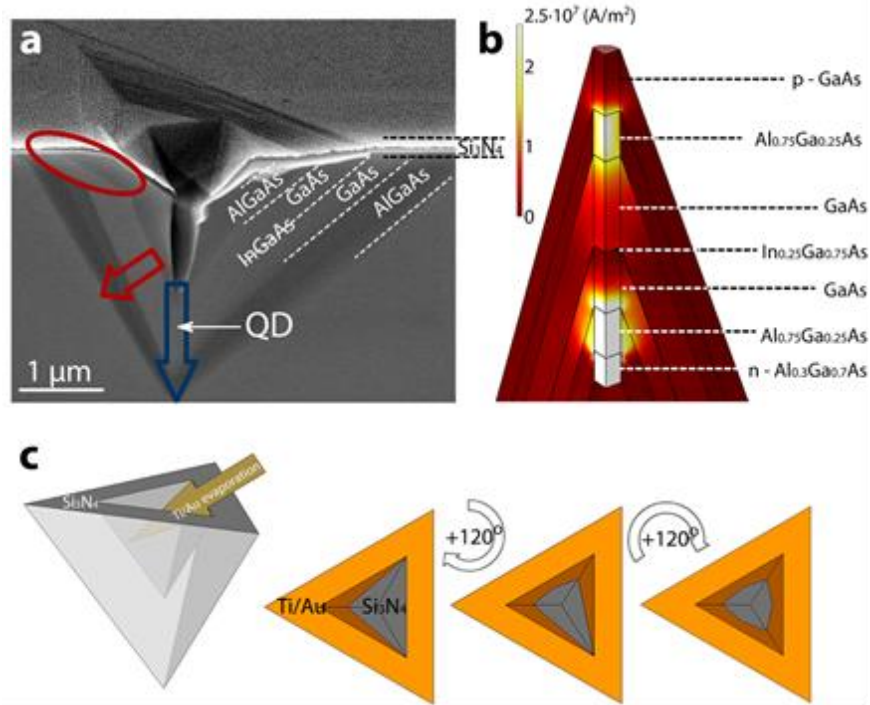

**Figure 2. Selective injection scheme and its realisation. a**, SEM image of cross-section view of a epitaxial layer structures grown within a pyramidal recess – the device is cleaved to half and still contained within the GaAs substrate in the initial, apex-down, geometry. The layer contrast is mostly due to different alloy compositions. A blue (red) arrow shows the preferential (most probable) direction of carrier injection as per discussion in our text. The region with irregular growth facets close to the surface, which potentially can lead to current leakage/shortcut, is encircled. The top of the sample is deposited with $Si_3N_4$ except the central part of the pyramid where the back contact of a µLED is formed. **b**, A colour-map representing current density in a simulated pyramidal structure (see Methods). The current density is concentrated in the regions of the vertical quantum wires (VQWRs). **c**, Processing steps to form an open aperture within $Si_3N_4$ at the central part of the pyramidal recess. Ti/Au (3/15 nm) is evaporated with an angle so that the edge of the recess would be partially shadowing gold deposition. By rotating the sample 120° in-plane twice more and repeating evaporation, an open $Si_3N_4$ area is left, which is etched away with $CF_4$ plasma exploiting gold as a protecting mask.

## Electroluminescence

Once the section of the wafer patterned with pyramidal recesses had been fully processed, all pyramids were contacted by evaporating a thin semitransparent layer of titanium (1 nm) and gold (20 nm), and bonding a Kapton insulated wire to the whole structure. At this stage, for simplicity, all µLEDs are contacted, and all potentially turn on. An example (Sample A – see

Methods and Supplementary Material for a detailed sample description) of gradually turning-on all (~1300) contacted μLEDs at 10 K is shown in Fig. 3c. The bright spots of spectrally unfiltered integrated electroluminescence match the initial pyramidal recess pattern demonstrating that each individual pyramid turns on without significant leakage through the GaAs substrate. This is confirmed by a macro-electroluminescence spectrum in Fig. 3b, where the dominant luminescence features at high injection current of 4 mA are InGaAs lateral quantum wires and lateral quantum wells clearly exceeding the QD luminescence and easily observable using a regular CCD imaging camera. Non-uniformity of intensity regions reflects differences between individual μLEDs, as each of them turn-on at a slightly different voltage, and the I-V curve shown in Fig. 3a is rather the characteristic of the whole ensemble and not of a single device. However, we stress that obviously nothing impedes to be selective to individual or a specific subset of μLEDs in future developments to address or drive them independently, as this would not require any sophistication in the processing.

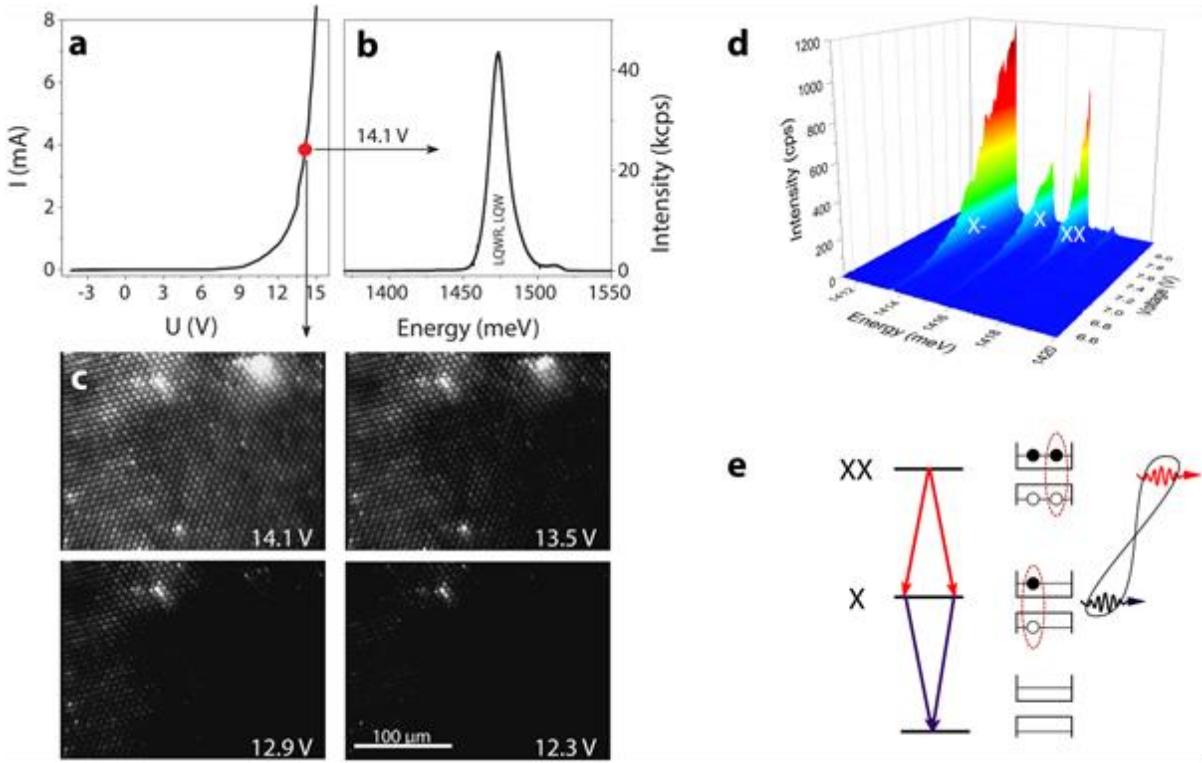

**Figure 3. Electroluminescence of μLEDs. a**, Current-voltage curve of the Sample A taken at 8 K. **b**, Macro electroluminescence spectrum taken under applied bias of 14.1 V. Dominant features are LQWRs and LQWs. **c**, Optical images of switching-on μLEDs with increasing applied bias. The light is not spectrally filtered. **d**, A single QD (Sample B1) electroluminescence intensity dependence on the applied voltage. Three dominant transitions are a negative trion (X-), exciton (X) and biexciton (XX). **e**, Polarization-entanglement realization scheme with XX and X transitions. The biexciton is composed of two electrons and two holes (two excitons) and is described by a singlet-like state. During the recombination cascade through the intermediate exciton state, a pair of polarization-entangled photons is emitted.

Several phenomena are possibly contributing to some inhomogeneity of electric injection properties: (1) an inhomogeneous etching profile during the BE step, (2) a complex non-planar surface profile of the back contact side due to slightly irregular MOVPE growth towards the centre of the pyramid, which tends to close the recess irregularly creating variable conditions for the back-contact formation, and (3) by the presence of resistance at the contact side (most probably the p-doped one). These phenomena somehow suggest why the turn-on voltage is unexpectedly high, around 7 V, as shown by a three-dimensional colour map (Fig. 3d) of a representative QD (Sample B1) electroluminescence dependence on the driving voltage, where the dominant transitions are a negative trion (X-), exciton (X) and biexciton (XX). We estimate overall extraction efficiency from these structures with a present design to be ~ 1% (see Methods for more details).

The average linewidth of exciton and biexciton transitions was found to be 138±34 µeV and 97±23 µeV, respectively. While practical applications will require, ideally, transform-limited linewidth photons, we argue that the broadening is not a fundamental issue here. The origin of it is mostly related to a "charged" vicinity of QDs inducing spectral wandering[24]. We observe that, at this scale of broadening, the external electric field which runs the device has no substantial effect. Ones of the clearly demonstrated sources of charge noise in this QD system are deep hole trapping levels in barriers[18]. The origin of the charge trapping states is most likely related to the processing-induced defects and MOVPE reactor state. We would like to stress that this can be overcome, and that we have a number of successfully fabricated samples with a high density of QDs with resolution-limited (<18 µeV) linewidth transitions under optical excitation[25]. As we will show later, spectral wandering, being a relatively slow process comparing to the timescale of the biexciton-exciton recombination cascade, does not preclude high fidelity of entanglement.

## Polarization-entangled photon emission

A QD initially populated by the electronic biexciton state (two electrons and two holes, a "singlet-like" configuration) fully recombines through the intermediate state exciton (one electron and one hole) emitting two successive photons which are polarization-entangled (Fig. 3e)[26,27]. Provided that the recombination process proceeds coherently, and a QD has symmetric carrier confinement potential, the emitted pair of photons is described by the Bell's state $|\psi\rangle = \frac{1}{\sqrt{2}}(|H_{xx} H_x\rangle + |V_{xx} V_x\rangle)$, where $|H\rangle$ and $|V\rangle$ are horizontal and vertical polarization states. However, the QD symmetry condition is very rarely fulfilled in general[28,29,30] – electron-hole exchange interaction lifts degeneracy in the exciton level by the energy amount called the fine-structure splitting (FSS), which causes beating (coherent oscillations) of the exciton spin state[18,31,32], later transferred to the two-photon polarization state. Integration over many photon-pair emission-detection events degrades or completely eliminates entanglement if the events are not temporally resolved. Reduction of FSS to the needed submicroelectronvolt level can be achieved either through various external tuning strategies[31,33,34,35] or exploiting intrinsic growth properties[11,36], like in this work.

Fig. 4a shows a map of a randomly selected area to measure the density of functional µLEDs with bright X and XX transitions and distribution of the FSS values, where filled-in triangles represent emitting µLEDs. Dysfunctional QDs represented by open triangles were either strongly charged or non-emitting due to possible defects in the vicinity of a QD. The numbers inside triangles are measured FSS values. Most of them are smaller than 4 µeV – a good agreement between previously found values from optically excited samples[18]. The full distribution of FSS values obtained from 94 µLEDs is shown in Fig. 4b with an average value of 2.9±1.8 µeV, while the standard deviation of the exciton transition energy is 2.6 meV.

Two µLEDs with FSS values of 0.7 and 0.4 µeV (marked by a red border in the map in Fig. 4a), and, importantly, without strong background coming from adjacent device electroluminescence, were selected to test for polarization-entanglement. Six polarization-resolved biexciton-exciton continuous-wave injection (CW) intensity correlation curves measured in rectilinear, diagonal and circular bases are shown in Fig. 4c from a device with 0.4±0.8 µeV FSS. Clearly obtained correlations between co-polarized biexciton and exciton photons in rectilinear and diagonal, and anticorrelation in circular bases are expected for polarization-entangled photons with a state $\frac{1}{\sqrt{2}}(|HH\rangle + |VV\rangle)$ which is equivalent to $\frac{1}{\sqrt{2}}(|DD\rangle + |AA\rangle) = \frac{1}{\sqrt{2}}(|RL\rangle + |LR\rangle)$, where $|R\rangle = \frac{1}{\sqrt{2}}(|H\rangle + i|V\rangle)$, $|L\rangle = \frac{1}{\sqrt{2}}(|H\rangle - i|V\rangle)$, $|D\rangle = \frac{1}{\sqrt{2}}(|H\rangle + |V\rangle)$, $|A\rangle = \frac{1}{\sqrt{2}}(|H\rangle - |V\rangle)$ are right/left-hand circular, diagonal and antidiagonal polarization states, respectively. The calculated fidelity curve (see Methods) is shown in the bottom graph of Fig. 4c. By selecting correlation events around 0 ns delay value from the time window of 0.5 ns, the fidelity value was found to be 0.73±0.06 exceeding the classical limit of 0.5. The fidelity to the expected maximally entangled state of photons emitted from a nearly adjacent µLED QD with FSS of 0.7±0.5 µeV was found to be 0.69±0.06.

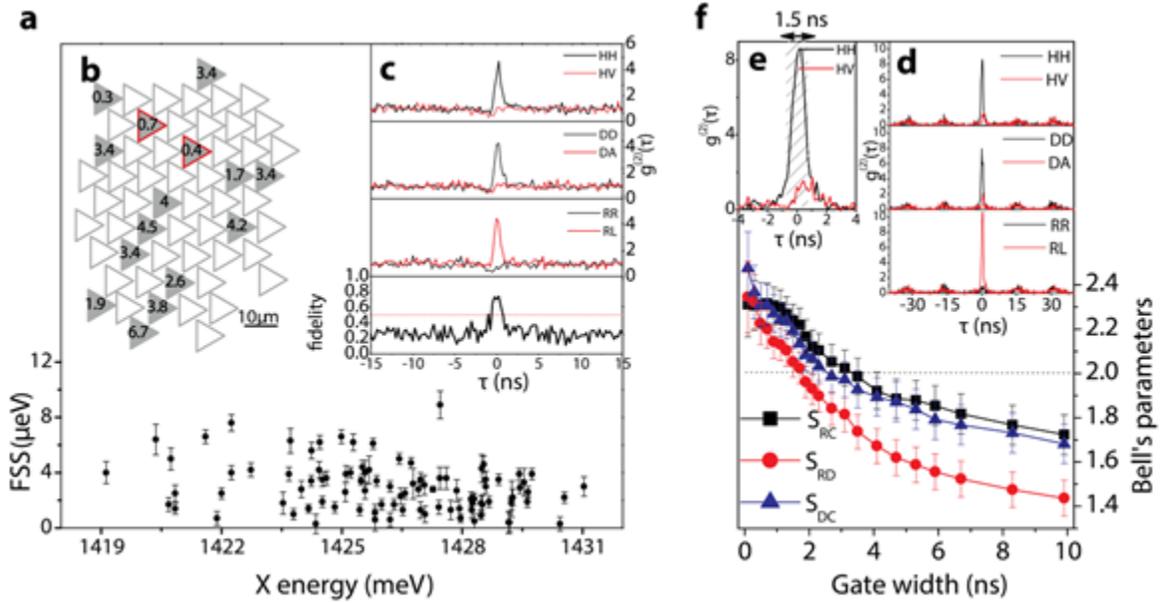

**Figure 4. Two-photon polarization state entanglement measurements. a**, Distribution of exciton fine-structure splitting values obtained from Sample B2. The error bars represent standard deviation of a fitted sinusoid curve, as described in Methods. **b**, A map of µLEDs from a randomly selected area. Filled-in triangles represent working devices with unquestionable exciton-biexciton emission. The numbers inside the triangles are the FSS values. Two devices framed in red and with a FSS equal to 0.7 and 0.4 µeV emitted polarization-entangled photons. **c**, CW measurement results of fidelity to the expected maximally entangled state obtained from a device with a FSS value of 0.4 µeV. The classical limit of 0.5 is significantly exceeded around 0 ns delay. **d**, Polarization-resolved second order correlation curves taken in rectilinear, diagonal and circular polarization bases under pulsed µLED injection. **f**, 3 Bell's parameters calculated by a simplified procedure using the results shown in **e**. By selecting correlation events from a time-window of 1.5 ns and preserving 75% of the initial intensity, all 3 parameters violated the classical limit of 2, proving non-classical nature of the polarization-state. The error propagation to the values of all parameters has been estimated using Poissonian uncertainty.

Practical applications will require a triggered source of photons, where ideally a single pair of polarization-entangled photons would be generated upon request. Therefore Sample B1 was injected with pulsed current with a 63 MHz repetition rate (see Methods). Polarization-resolved second-order correlation curves obtained from a QD with a FSS of 0.2±0.2 µeV are shown in Fig. 4d. Expected correlations and degree of entanglement quantified by the fidelity value $f=$ 0.678±0.023 confirmed the non-classical state of the photons. While the FSS value of this representative QD is small enough to have a substantial effect on entanglement degradation, a number of other phenomena are contributing. Among them are (1) a finite pulse-width of the current (1.4 ns) which partially causes repopulation of a QD and (2) a small background emission coming from the adjacent µLEDs. By using a time-gating technique, which could be defined as a selection of correlation events from a specific time window, the contribution of these phenomena can be significantly reduced. For example, by selecting events from a 1.5 ns window (shown in the inset of Fig. 4e) and preserving 75% of the total two-photon detection events, the fidelity increased to 0.823±0.019 (with a shorter time window of 0.1 ns, fidelity goes as high as 0.881±0.042, but care needs to be taken, as the photon statistics is also decreased). The taken correlation measurements allow a simplified estimation of three Bell parameters[37] used in quantum communication protocols[38]. By selecting correlation events from 1.5 ns window, the parameters were measured to be $S_{RD}$=2.053±0.070, $S_{DC}$=2.191±0.075 and $S_{RC}$=2.239±0.074, all violating Bell's inequalities.

## Conclusions and outlook

In summary, we have introduced a quantum photonics technology which potentially enables fabrication of site-controlled, scalable arrays of electrically driven sources of polarization-entangled photons with high entanglement quality. Compatibility with semiconductor fabrication technology, good reproducibility and control of the position make these devices attractive candidates for integrated photonic circuits for quantum information processing.

Nevertheless, work is needed to further improve our sources to achieve full capability in the contest of quantum information processing: achieving photon indistinguishability across all devices, external control of the remaining fine-structure splitting, and enhancing photon extraction efficiency are among the top priorities. Strain with a possible combination of electric field[12,39] is one of the most promising tuning strategies of the emission energy and fine-structure splitting, which, we anticipate, already proved to be efficient within the first prototypes based on the Pyramidal QD system. The typically observed emission energy distribution (standard deviation of 2.5-3 meV) as in our sample(s) could be easily corrected by these methods, delivering the same excitonic emission energies for each Pyramid. Our short-term ambition will be dedicated to improving source tuneability, for example, by implementing a six-legged semiconductor-piezoelectric device, as in Ref.40, also to achieve full control of the FSS associated to each Pyramid. Another critical parameter, the transition linewidth, which ideally is expected to be transform-limited, is indeed subject to the electric (and magnetic) fields present in the vicinity of a QD[24], including the external electric field applied for device operation. However, this is not a fundamental limitation, as it was already demonstrated that electrically driven QD devices can emit photons with a linewidth close to the theoretical limit[41]. By optimizing growth, device and interface design and processing conditions to minimize defect density and charge accumulation, high spectral purity is expected to be achieved. Indeed, a linewidth of a few µeV (resolution limited) from non-resonantly optically excited Pyramidal QDs was already demonstrated[25]. Finally, photon collection efficiency can be increased with proper on-chip lensing and waveguiding strategies[42,43,44], which potentially allow photon extraction efficiency up to 80%.

## Methods
**Sample growth.** All the presented results were obtained from three samples referred in the text as Sample A, B1 and B2. The structures were grown by metalorganic vapour phase epitaxy (MOVPE) on (111)B oriented GaAs substrates pre-patterned with tetrahedron recesses with a pitch of 7.5 µm for the sample A, and 10 µm for the samples B1 and B2. Among them, the ones with µLEDs emitting entangled photons are samples B1 and B2, which were from an identical epitaxial structure but processed in different runs. The nominal QD composition and thickness of latter samples are $In_{0.25}Ga_{0.75}As$ and 0.55 nm. The full epitaxial structure, doping and the role of each layer are given in the Supplementary Material.

**Micro-LED fabrication.** Detailed steps of µLED fabrication are given in Supplementary Material.

**Measurements.** Measurements were taken at 10 K using a helium closed-cycle cryostat. Electroluminescence was collected in a standard micro-photoluminescence set-up, using 100x magnification, a 0.80 NA long-working distance objective, which enabled probing a single device at a time.

The fine-structure splitting measurements were taken by using a combined multiple measurement and fitting procedure[37]. Linear polarization components were analysed by placing a polarizer in front of the monochromator and rotating a half-wave plate with a step of 1.5 deg. Exciton and biexciton transitions were fitted with Lorentzian fits; the corresponding peak centres were subtracted and the resulting data fitted by a sinusoid curve, where its amplitude is equal to the FSS value (an example is given in Supplementary Material). The standard deviation was taken as an error.

Polarization-entanglement was measured by discriminating exciton and biexciton transitions with two monochromators acting as narrow band-pass filters. Polarization projections were selected by an appropriate orientation of half-(quarter-)wave plates to the respect of polarizing beam-splitters placed after the monochromators equipped with 950 grooves/mm gratings (TE/TM diffracted intensity ratio ~1 at 877 nm). Fiber-coupled avalanche-photo diodes were used to detect intensity at a single photon level. In such configuration, two polarization-resolved, second-order correlation curves were measured simultaneously. The error propagation to $g^{(2)}$, fidelity values has been estimated by using a Poissonian uncertainty.

In pulsed excitation mode, samples were injected by pulses shaped with a positive DC offset set slightly below the µLED injection threshold (typically a few volts), with superimposed pulses of 1.4 ns width, reaching maximum voltage as high as 20 V and repetition rate between 63 MHz and 80 MHz. A typical polarization-entangled photon pair detection rate in these experimental conditions was ~1000 pairs/hour, with overall estimated extraction efficiency from our structures ~ 1%. The final counts are also affected by the ~1.4% efficiency of our µPL set-up and a broad non-Gaussian µLED emission profile.

**Fidelity and Bell parameters calculation.** Typically the two-photon polarization state can be estimated by a quantum state tomography procedure [45] which allows reconstruction of a density matrix $\rho$ from a set of 16 intensity measurements. Since the expected maximally entangled state $|\psi\rangle = \frac{1}{\sqrt{2}}(|HH\rangle + |VV\rangle)$ is known, the procedure can be simplified by reducing the number of measurements, which allows obtaining only the density matrix elements necessary to calculate the fidelity $f = \langle \psi | \rho | \psi \rangle$ of the entangled state $|\psi\rangle$ : $f = \frac{1}{4}(1 + C_R + C_D - C_C)$, where $C_R$, $C_D$ and $C_C$ are degrees of correlations taken in rectilinear, diagonal and circular polarization bases[31,46]. The degree of correlation is defined as $C_{basis} = (g^{(2)}_{xx,x} - g^{(2)}_{xx,\bar{x}})/(g^{(2)}_{xx,x} + g^{(2)}_{xx,\bar{x}})$, where $g^{(2)}_{xx,x(\bar{x})}$ is the second-order correlation function with $xx(x)$ being polarization of a biexciton (exciton) and $\bar{x}$ being orthogonal polarization of an exciton. Fidelity value greater than 0.5 is a quick indicator of entanglement.

The degrees of correlation $C_R$, $C_D$ and $C_C$ are used to calculate the simplified Bell parameters[37]. Without the aim to rule-out local hidden-variable theories, the traditional CHSH form[47] of inequality obtained from the measurements with four different combinations of polarizers, can be simplified and expressed as $S_{RD} = \sqrt{2}(C_R + C_D) \leq 2$. Two more different Bell parameters are calculated and known as $S_{DC} = \sqrt{2}(C_D - C_C) \leq 2$ and $S_{RC} = \sqrt{2}(C_R - C_C) \leq 2$.

**Current density simulations.** Finite element method simulations were performed using COMSOL Multiphysics 5.0. The simulations were obtained by solving the Poisson's equation in conjunction with the continuity equations in order to calculate the voltage and carrier density (electrons and holes) in a 2D geometry (described below). Rotational symmetry along the vertical direction was applied to confer a 3D structure to the simulated geometry and improve the approximation to the real pyramidal structure. The 2D geometry consisted of half cross section of a regular triangular pyramid with side length of 7 µm cut along the centre of a face. The vertical wire in the centre of the 3D geometry was approximated to a cylinder with diameter of 100 nm. The simulated structure has 7 internal layers which are (starting from the top layer in Fig. 1d) : GaAs with thickness 60 nm, $Al_xGa_{1-x}As$ with x=0.75 and thickness 45 nm, GaAs with thickness 90 nm, $In_{0.25}Ga_{0.75}As$ with thickness 1 nm, GaAs with thickness 60 nm, $Al_xGa_{1-x}As$ with x=0.75 and thickness 45 nm and finally $Al_xGa_{1-x}As$ with x=0.3 and thickness 30 nm. In the vertical quantum wire the Al concentrations in the different layers were reduced to include the segregation effects[19]: x=0.3 were substituted with x=0.05, and x=0.75 were substituted with x=0.26. The top layer was p-doped with concentration $1 \cdot 10^{18}$ cm$^{-3}$ and the external surface was set at constant voltage of 1.5 V. The bottom layer was n-doped with concentration $1 \cdot 10^{18}$ cm$^{-3}$ and set at ground. The current density values reported in Fig. 1d are the nodal value calculated from the electric field distribution *E* and the conductivity *σ* using the relation *J=σE*.

## Acknowledgements

This research was enabled by the Irish Higher Education Authority Program for Research in Third Level Institutions (2007-2011) via the INSPIRE (Integrated NanoScience Platform for Ireland) programme, and by Science Foundation Ireland under grants 10/IN.1/I3000 and 07/SRC/I1173. The authors are grateful to Dr K. Thomas for the MOVPE system support.

## Author contributions
T.H.C. fabricated the devices. G.J. and S.T.M. carried out optical characterization, data processing and analysis. A.P. made theoretical calculations. A.G. contributed to sample growth and MOVPE system operation. E.P. conceived the study, participated in its design and coordination, and equally contributed to writing the manuscript with G.J. All authors commented on the final manuscript.

## Competing financial interests
The authors declare no competing financial interests.

## Materials & Correspondence
Correspondence and requests for materials should be addressed to gediminas.juska@tyndall.ie